\documentclass[
reprint,
amsmath,amssymb,
aps,
natbib=true,
nofootinbib
]{revtex4-2}

\usepackage{graphicx}% Include figure files
\usepackage{dcolumn}% Align table columns on decimal point
\usepackage{bm}% bold math
\usepackage{xcolor}
\usepackage[colorlinks=true,
linkcolor=blue,
citecolor=blue,
urlcolor=blue, pdfborder={0 0 1}]{hyperref}

\usepackage{orcidlink}

\begin{document}

\preprint{APS/123-QED}

\title{Swarmalators with frequency-weighted interactions}% Force line breaks with \\
%\thanks{A footnote to the article title}%

\author{R. Senthamizhan\,\orcidlink{0009-0000-0669-3336}}
 %\altaffiliation[Also at ]{Physics Department, XYZ University.}%Lines break automatically or can be forced with \\
\author{R. Gopal\,\orcidlink{0000-0001-5627-9758}}%
\email{Corresponding author: gopalphysics@gmail.com}
\author{V.K. Chandrasekar\,\orcidlink{0000-0002-2220-9310}}%
 \email{Corresponding author: chandru25nld@gmail.com}
\affiliation{%
	Department of Physics, Centre for Nonlinear Science and Engineering, 
	School of Electrical and Electronics Engineering,
	\href{https://ror.org/032jk8892}{SASTRA Deemed University}, 
	Thanjavur 613 401, India
}%

%\date{}% It is always \today, today,
             %  but any date may be explicitly specified

\begin{abstract}
We investigate the role of frequency-weighted interactions in a solvable model of one-dimensional (1D) swarmalators confined to a ring, where both spatial and phase couplings are scaled by the heterogeneous natural frequencies of individual agents. Our analysis identifies three distinct collective states: the asynchronous state , the phase-wave state , and the bistrip mixed state characterized by antipodal clusters that are internally split into frequency-dependent sub-strips. We further establish that the onset of abrupt transitions are driven by heterogeneous coupling. Using a self-consistency analysis, we precisely determine the conditions for dynamical transitions among the identified states, thereby extending the theoretical understanding of swarmalator dynamics under heterogeneous interaction rules, which are in good agreement with the numerical simulation results.
\begin{description}
\item[Key words]
1D swarmalators, Frequency weighted coupling
 
\end{description}
\end{abstract}

%\keywords{Suggested keywords}%Use showkeys class option if keyword
                              %display desired
\maketitle

%\tableofcontents

\section{Introduction}
\label{sec:intro}
Swarmalators are a class of oscillators that can both synchronize their internal phases and form spatial patterns simultaneously. They combine the dynamics of phase oscillators, such as those in the Kuramoto model \cite{kuramoto2005self}, with self-propelled particle systems like the Vicsek model \cite{vicsek1995novel}. While these two aspects have often been studied separately under various configurations \cite{childs2008stability, manoranjani2022quenching,manoranjani2024asymmetric,  wang2025higher, horton2025order, dutta2025stability, gu2025mixing, jin2025role}, many natural and engineered systems ranging from bacterial colonies \cite{belovs2017synchronized} and chemical nanomotors \cite{zhou2020coordinating} to robotic swarms \cite{trianni2009self} exhibit a fundamental interplay between synchronization and swarming. This makes the swarmalator framework a natural setting for exploring how swarming and synchronization interact.

Tanaka \cite{tanaka2007general} and Iwasa \cite{iwasa2010dimensionality} laid the foundation for the mobile chemotactic oscillators, in which individual units move in response to a chemical stimulus in a dynamic and coupled manner. The chemotactic mobile oscillators demonstrated spatial clustering, phase synchronization, and their interplay. Building on this concept,  the swarmalator model was first introduced by O’Keeffe et al. in 2017 \cite{okeeffe_2017} and has since attracted significant attention due to its dynamical richness. Following this foundation, several studies have explored swarmalator systems under different model configurations and interaction rules \cite{lee2021collective, ceron2023diverse, xu2024collective, senthamizhan2024data, yadav2024exotic, yadav2025collective}. Although two- and three-dimensional swarmalators are more realistic and have been realized in various applications \cite{lucas2024interactive, toiviainen2025modeling}, analitical progress remains limited. Only a sparse amount of studies are reported and a rigorous analytical framework is still elusive due to the presence of long-range, short-range interaction kernels and their chaotic nature \cite{ansarinasab2024spatial}. To overcome these difficulties, a one-dimensional version of the model was introduced by O’Keeffe et al. in 2022 \cite{keeffe_2022_ring}, where agents are confined to a ring. This one-dimensional toy model is particularly valuable because it is analytically tractable, making it a powerful tool for gaining theoretical insight into the underlying dynamics. Building on this solvable framework, a number of works have been extended this 1D model to include effects such as thermal noise, time delay, periodic forcing, and both uncorrelated and correlated pinning \cite{hong2023swarmalators, blum2024swarmalators, anwar2024forced, sar2023pinning, sar2023swarmalators}. Even solvable two-dimensional extensions have also been investigated \cite{o2024solvable}. These developments show that modifying the interaction rules will lead to new dynamical behaviors while still allowing analytic or semi-analytic treatment.

Motivated by this line of work, we study a specific instance of a solvable swarmalator model, where the coupling interactions are weighted by the agents’ intrinsic drives. This form of heterogeneous coupling has been extensively explored in the Kuramoto literature \cite{xu2016synchronization, xu2016dynamics, bi2017nontrivial, ameli2024two, zou2024solvable}. In those studies, scaling the coupling strength with natural frequencies has been shown to produce explosive synchronization transitions and various complex dynamical states, including non-trivial standing waves, Bellerophon states, traveling waves, and chimera-like states. Frequency-weighted coupling is especially interesting because it is physically motivated by systems where an agent’s ``activity’’ or ``energy’’ dictates its influence on the collective \cite{kumar2015experimental, leyva2012explosive}. Extending this idea to swarmalators provides a natural way to test how such heterogeneous influences shape the interplay between synchronization and swarming. In this work, we employ a combination of numerical, self-consistency and semi-analytic analysis to identify the dynamical states and critical boundaries among them, thereby providing a comprehensive framework for investigating the generalized frequency weighted 1D swarmalator model. This alternative perspective offers fresh insight into the role of heterogeneity in synchronization-swarming interactions and distinguish our work from previous studies in both the swarmalator and frequency-weighted coupling literature.
\begin{figure*}[t]
	\hspace*{-0.7cm}
	\includegraphics[width=15cm]{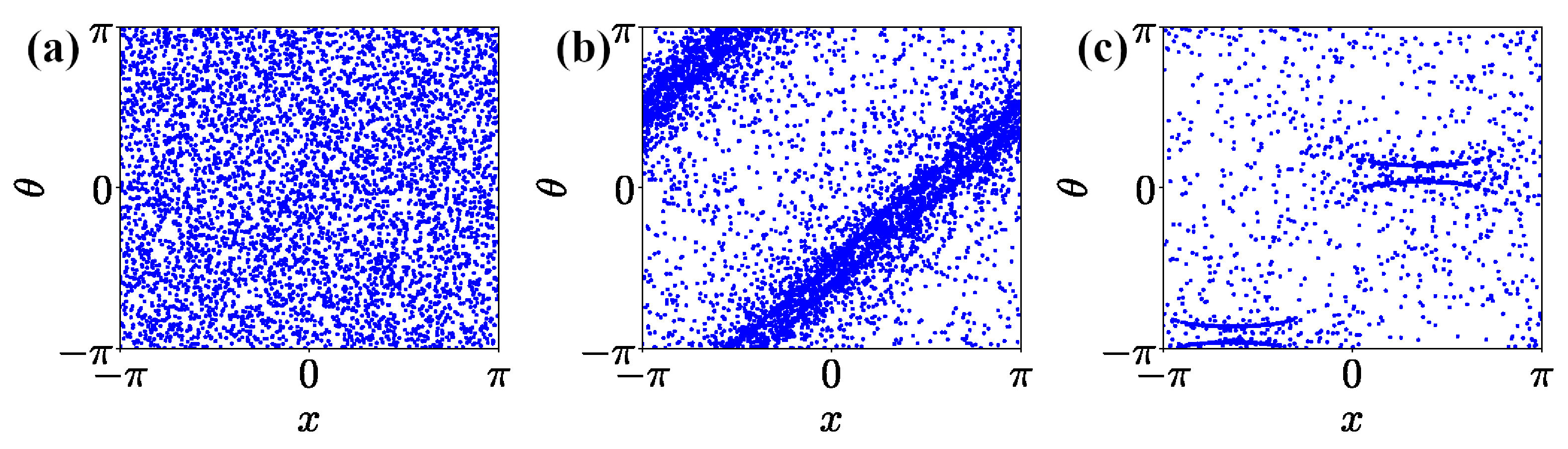} % Replace with your actual figure file
	\caption{Phase space portraits ($x$ vs $\theta$) of the three primary states observed for a fixed spatial coupling $J=12$. (a) The disordered \textbf{AS} at $K=-8$, (b)  \textbf{PW} at $K=0.5$, and (c)  \textbf{BM} at $K=10$.}
	\label{fig:1}
\end{figure*}

The paper is organized as follows. We begin with the introduction in \textcolor{blue}{Sec.} \ref{sec:intro} and then provide a brief description of the model in \textcolor{blue}{Sec.} \ref{sec:model}. The results and analyses in \textcolor{blue}{Sec.} \ref{sec:results} present the numerical exploration of the observed states, their properties, and the corresponding transitions, followed by the self-consistency analyses that connects the numerical findings with analytical and semi-analytical results. Finally, \textcolor{blue}{Sec.} \ref{sec:discussion} provides concluding remarks and outline future directions.

\section{MODEL DESCRIPTION}
\label{sec:model} 
We consider a system of $N$ swarmalators moving on a one-dimensional ring. The state of each agent $i$ is described by its spatial position $x_i \in [-\pi, \pi]$ and its internal oscillatory phase $\theta_i \in [-\pi, \pi]$. The dynamics are governed by the following coupled differential equations:
\begin{align}
	\dot{x}_i &= v_i + \frac{J}{N} \sum_{j=1}^{N} f_1(x_j, x_i) g_1(\theta_j, \theta_i) \label{eq:x_dot}, \\
	\dot{\theta}_i &= \omega_i + \frac{K}{N} \sum_{j=1}^{N} f_2(x_j, x_i) g_2(\theta_j, \theta_i). \label{eq:theta_dot}
\end{align}

Here, $v_i$ and $\omega_i$ are the intrinsic velocity and natural frequency of agent $i$, respectively. Both quantities are drawn from a Cauchy distribution, defined by the probability density function:
\begin{align}
    g(\omega_i) = \frac{1}{\pi \gamma \left[ 1 + \left( \frac{\omega_i - \omega_0}{\gamma} \right)^2 \right]}.
\end{align}
The parameters $J$ and $K$ represent the base strengths of the spatial and phase coupling. The functions $f_1 = \sin(x_j - x_i)$, $f_2 = \cos(x_j - x_i)$, $g_1 = |\omega_i| \cos(\theta_j - \theta_i)$ and $g_2 = |\omega_i| \sin(\theta_j - \theta_i)$, defines the interaction among swarmalators. The forms of $g_1$ and $g_2$ imply that agents with higher intrinsic frequencies have a proportionally stronger influence on their neighbors, both spatially and in phase. Conversely, agents with frequencies near zero have a negligible effect on the collective dynamics. This introduces a non-trivial form of quenched disorder \cite{hong2022first} where the heterogeneity is directly correlated with the agents' intrinsic dynamics. The subsequent sections explore the resulting collective dynamical states, their properties, and the transitions between them.

\section{RESULTS AND ANALYSES}
\label{sec:results} 
\subsection{Numerical Results}
Numerical simulations of Eqs.~(\ref{eq:x_dot},\ref{eq:theta_dot}) were carried out using an adaptive RK45 solver with $N=5000$ swarmalators and $T=50,000$ time-steps. The intrinsic velocities and frequencies $(v_i,\omega_i)$ were drawn from Cauchy distributions with central frequencies $v_0=\omega_0=0$ and scale parameter $\gamma=1$. The simulations reveal three distinct dynamical states depending on the coupling strengths $J$ and $K$. The asynchronous state (AS) appears for $J=12, K=-8$ (Fig.~\ref{fig:1}(a)), where the swarmalators are uniformly distributed in the $(x,\theta)$ plane, occupying all values without correlation. The phase-wave state (PW) occurs for $J=12, K=0.5$ (Fig.~\ref{fig:1}(b)), where the swarmalators align along a diagonal strip in the $(x,\theta)$ plane, showing a near-linear correlation $x \approx \theta$. 

\begin{figure}[h!]
	\hspace*{-0.5cm}
	\includegraphics[width=7.0cm]{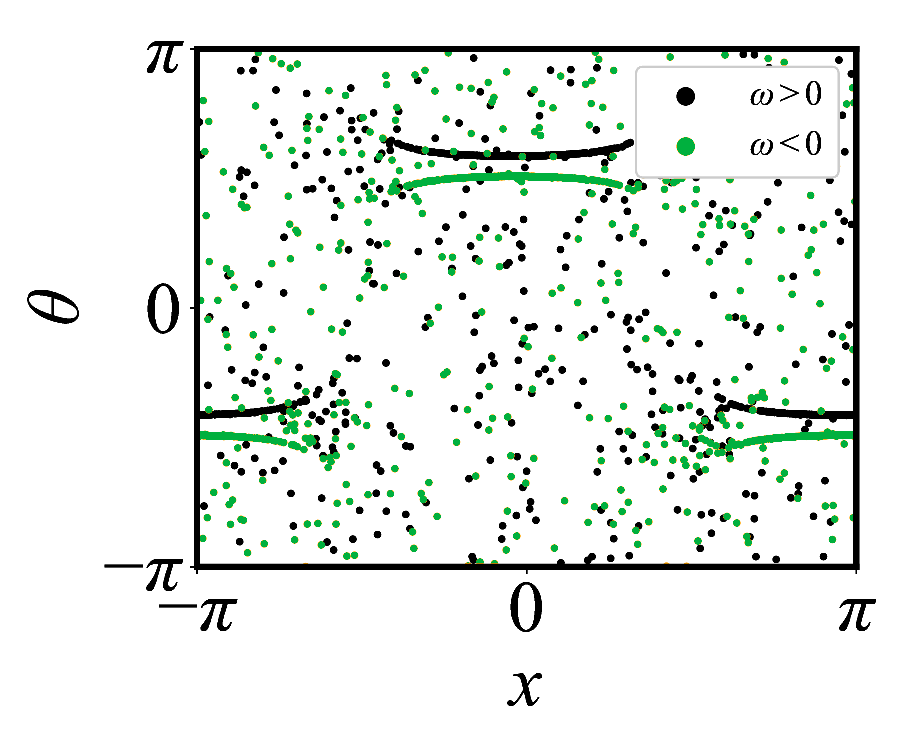}
	\caption{Subclusters of the BM state for $J=12$, $K=10$ in the \((x,\theta)\) plane, with colors indicating the sign of its intrinsic frequency \(\omega\)}
	\label{fig:2}
\end{figure}

The bi–strip mixed (BM) state is obtained for $J=12, K=10$ (Fig.~\ref{fig:1}(c)). In this state, the swarmalators form two antipodal clusters, each of which splits into two parallel sub-strips determined by the sign of the natural frequency ($\omega>0$ or $\omega<0$), as illustrated in Fig.~\ref{fig:2}. The BM state arises due to the frequency-weighted coupling, where $|\omega|$ facilitates frequency based clustering.

To characterize the observed collective states, we employ the complex order parameters, designed to capture the coupled space-phase correlations inherent to swarmalators \cite{keeffe_2022_ring}:
\begin{align}
	W^{\pm} = S_{\pm} e^{i \psi_{\pm}} = \frac{1}{N} \sum_{j=1}^{N} e^{i(x_j \pm \theta_j)}.
	\label{cop}
\end{align}
The magnitudes, $S_+$ and $S_-$, quantify the degree of coherence in the sum and difference channel of the agents, respectively.

\begin{figure}[h]
	\hspace*{-0.7cm}
	\includegraphics[width=8cm]{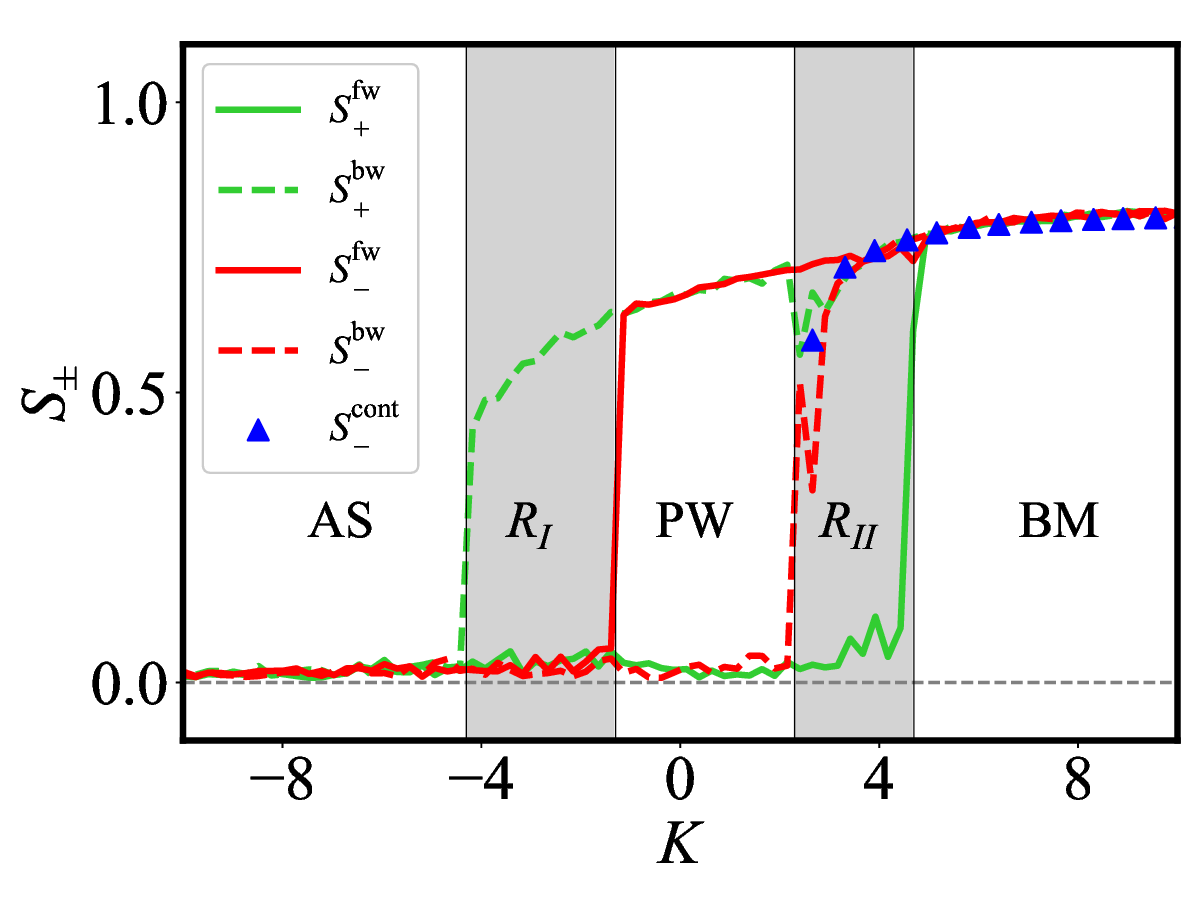} % Replace with your actual figure file
	\caption{Sweep of the coupling parameter $K$ at fixed $J=12$, showing the order parameters $S_{\pm}$. 
		The intermediate intervals $R_{I}$ and $R_{II}$ indicate bistable windows between AS–PW and PW–BM states. 
		Solid curves represent forward sweeps ($S^{\mathrm{fw}}_{+}$, $S^{\mathrm{fw}}_{-}$), while dashed curves denote backward sweeps ($S^{\mathrm{bw}}_{+}$, $S^{\mathrm{bw}}_{-}$). 
		The \textcolor{blue}{$\blacktriangle$} markers indicate the stable BM branch obtained from self-consistency analysis.}
	\label{fig:3}
\end{figure}

The order parameters $W^{\pm}$ provide a clear distinction between the dynamical states, as shown in Fig.~\ref{fig:3}. For $J=12$, the plot of $S_{\pm}$ versus $K$ reveals the sequence of states and their associated transitions. The plot shows both the forward and backward sweeps of the order parameters $S_{\pm}$, where the green and red solid curves represent forwardly swept $S_+$ and $S_-$ respectively. while, dashed curves of same color code represents backward sweep. In the forward direction, the AS state shows both $S_+$ and $S_-$ remain close to zero, transitioning to PW state at $K \approx -1.3$. In the PW state, the $S_-$ dominates with $S_- > 0$ and $S_+ \approx 0$ showing transition to BM state near $K \approx 4.7$. Whereas, in the BM state, both order parameters become nonzero, indicating a higher degree of synchrony compared with AS or PW. From the Figs. \ref{fig:1} (b) and \ref{fig:1} (c) we can observe that in both the PW and BM, there are swarmalators drifting outside the locked population contributing to the lower order parameter values $S_{\pm} \approx 0.8$. When the coupling strength $K$ is swept downward, the BM state loses stability at $K \approx 2.3$ and PW loses at $K \approx -4.3$ leading to two bistable windows: $R_I$, where AS and PW coexist, and $R_{II}$, where PW and BM coexist. 

\begin{figure}[h]
	\hspace*{-0.7cm}
	\includegraphics[width=8.5cm]{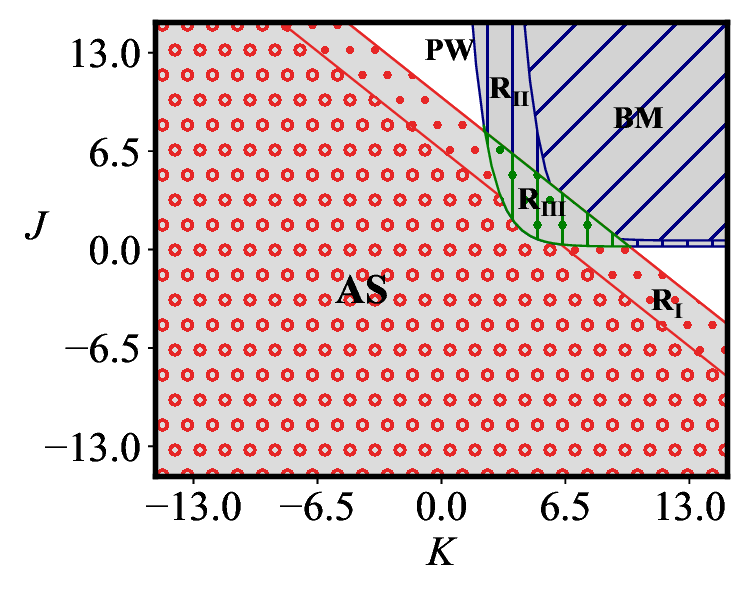}
	\caption{$J$–$K$ parameter space illustrating the full set of observed states. 
		The intermediate regions $R_{I}$ and $R_{II}$ denote bistable windows between AS–PW and PW–BM states, while $R_{III}$ represents a multistable regime in which all three states coexist.}
	\label{fig:4}
\end{figure}

We can also observe how these hysteretic transitions evolve, by sweeping both $J$ and $K$ leading to a global view of all the three dynamical states as shown in Fig.~\ref{fig:4}. The AS state spans all four quadrants, dominating weak and negative $K$, the BM state is confined to the $(+J,+K)$ quadrant as it requires strong positive coupling for coherence in both $S_{\pm}$, and the PW state appears in two wedge–shaped regions separating AS and BM, mediating between strong order and incoherence. The forward and backward sweeps reveal two bistable strips $R_I$ (AS-PW) and $R_{II}$ (PW-BM). In addition, the bistable regions $R_I$ and $R_{II}$, overlaps to give a multistable region $R_{III}$. To clearly understand the transitions and the associated bifurcations, we explore each of the observed states analytically in the following subsections.

\subsection{Self-consistency Analysis}
To simplify the analysis, the system Eqs. (\ref{eq:x_dot})-(\ref{eq:theta_dot}) must be rewritten in more convenient coordinates. Since the order parameters in Eq.~(\ref{cop}) capture the dynamics in the two channels $(x \pm \theta)$, these channels of synchronization can be more clearly exposed by introducing the sum–difference variables \cite{yoon2022sync} $(\epsilon,\eta)$ via
\begin{align}
	\epsilon_i &= x_i + \theta_i, \\
	\eta_i     &= x_i - \theta_i.
\end{align}

under which Eqs.~(\ref{eq:x_dot})–(\ref{eq:theta_dot}) become

\begin{align}
	\dot{\epsilon_i} &= v_i + \omega_i 
	+ |\omega_i|\Big[ J_+ S_+ \sin(\psi_+ - \epsilon_i) \notag \\
	&\qquad\quad + J_- S_- \sin(\psi_- - \eta_i) \Big] 
	\label{eq:epsilon_dot}, \\
	\dot{\eta_i} &= v_i - \omega_i 
	+ |\omega_i|\Big[ J_- S_+ \sin(\psi_+ - \epsilon_i) \notag \\
	&\qquad\quad + J_+ S_- \sin(\psi_- - \eta_i) \Big].
	\label{eq:eta_dot}
\end{align}
with $J_\pm = (J \pm K)/2$. Each complex mean field $W_\pm = S_\pm e^{i\psi_\pm}$ rotate at a constant collective frequency; we remove this trivial drift by working in frames rotating at $\Omega_\pm$:
\begin{align}
	\psi_\pm(t) &= \Omega_\pm t.
	\label{eq:collective_frequency}
\end{align}

Defining the deviations from these frames,
\begin{align}
	\phi_i &= \epsilon_i - \psi_+, \\
	\chi_i &= \eta_i - \psi_-.
	\label{eq:rotating_vars}
\end{align}

and the detunings
\begin{align}
	A_i &= (v_i + \omega_i) - \Omega_+, \nonumber \\
	B_i &= (v_i - \omega_i) - \Omega_-.
	\label{eq:detunings}
\end{align}

Using $\sin(\psi_+ - \epsilon_i) = -\sin\phi_i$ and $\sin(\psi_- - \eta_i) = -\sin\chi_i$, Eqs.~(\ref{eq:epsilon_dot})–(\ref{eq:eta_dot}) become
\begin{align}
	\dot{\phi}_i &= A_i - |\omega_i|\!\left(J_+ S_+ \sin\phi_i + J_- S_- \sin\chi_i\right),
	\label{eq:phi_dyn}\\
	\dot{\chi}_i &= B_i - |\omega_i|\!\left(J_- S_+ \sin\phi_i + J_+ S_- \sin\chi_i\right).
	\label{eq:chi_dyn}
\end{align}
The Eqs. (\ref{eq:phi_dyn}), (\ref{eq:chi_dyn}) are of Adler type with motion in a tilted washboard potential \cite{adler2006study}, with locking possible only if the effective tilt is smaller than the coupling amplitude. Eqs. (\ref{eq:phi_dyn}–\ref{eq:chi_dyn}) provide the foundation for subsequent analyses.

\subsection*{(i) Asynchronous state}
The AS is characterized by a nearly uniform distribution of swarmalators over both $x$ and $\theta$ (Fig.~\ref{fig:1} (a)), resulting in vanishing order parameters $S_\pm \approx 0$ and, hence, no position–phase correlation. AS typically shares a stability boundary with the PW state. We determine the onset of this instability by a self-consistency analysis. Since in the thermodynamic limit $N \to \infty$ the order parameters are determined by ensemble averages, and the index $i$ can be dropped in favor of continuous variables.

\medskip
\noindent

\medskip
To probe the linear PW onset, set $S_- = 0$ and let $S_+ \ll 1$. Then (\ref{eq:phi_dyn}) decouples to
\begin{align}
	\dot{\phi} &= A - |\omega|\,J_+ S_+ \sin\phi.
	\label{eq:phi_async}
\end{align}

Locked oscillators satisfy $\dot{\phi}=0$, i.e.
\begin{align}
	\sin\phi^* &= \frac{A}{|\omega|\,J_+ S_+}, 
	\cos\phi^* &= \sqrt{1-\!\left(\frac{A}{|\omega|\,J_+ S_+}\right)^{\!2}}.
	\label{eq:lock_condition}
\end{align}

with locking condition $|A|\le |\omega|\,J_+ S_+$. At the onset $S_+\to0^+$ this defines a locked band $A\in[-|\omega|J_+S_+,\,|\omega|J_+S_+]$.

The self-consistency integrals are defined as,

\begin{align}
	S_\pm = \iint g_v(v) g_\omega(\omega) e^{i(x \pm \theta)} dv d\omega.
	\label{eq:self_consistency_general}
\end{align}

by averaging the locked contribution \cite{xu2016synchronization}:
\begin{align}
	S_+ &= \iint_{\text{locked}} g_v(v)\,g_\omega(\omega)\,\cos\phi^*\,\mathrm{d}v\,\mathrm{d}\omega.
	\label{eq:self_consistency}
\end{align}

with $\phi^*$ being the locked solution. 
By changing the variables with the help of $A = (v + \omega) - \Omega_+$ and, by symmetry of $g_v$ and $g_\omega$ about zero, set $\Omega_+=0$ at onset. For small $S_+$ we take $g_v(A-\omega)\approx g_v(-\omega)$. The inner integral evaluates exactly as:
\begin{align}
	S_+ &\approx \int_{-\infty}^{\infty} g_\omega(\omega)\,g_v(-\omega)\!
	\left[ \int_{-|\,\omega|J_+S_+}^{|\,\omega|J_+S_+}
	\sqrt{1-\!\left(\frac{A}{|\omega|J_+S_+}\right)^{\!2}}\,
	\mathrm{d}A \right]\mathrm{d}\omega \notag\\[2pt]
	&= \int_{-\infty}^{\infty} g_\omega(\omega)\,g_v(-\omega)\,
	\left( \frac{\pi}{2}\,|\omega|\,J_+ S_+\right)\mathrm{d}\omega.
	\label{eq:inner_done}
\end{align}

\medskip

For unit-width Cauchy case, $g_\omega(\omega)=\big[\pi(1+\omega^2)\big]^{-1}$ and $g_v(v)=\big[\pi(1+v^2)\big]^{-1}$,
Eq.~(\ref{eq:inner_done}) becomes
\begin{align}
	S_+ &= \frac{J_+ S_+}{2\pi}
	\int_{-\infty}^{\infty}\frac{|\omega|}{(1+\omega^2)^2}\,\mathrm{d}\omega, \nonumber \\
	&= \frac{J_+ S_+}{2\pi}.
\end{align}

since $\int_{-\infty}^{\infty}\!|\omega|/(1+\omega^2)^2\,\mathrm{d}\omega=1$. Thus the AS loses stability to PW when
\begin{align}
	J_{+,c} &= 2\pi,
	\label{eq:Jplus_crit}
\end{align}

with $J_+=(J+K)/2$. By the same linear argument, the negative channel $\eta$ has the same threshold. $J_{+,c}$ only captures the loss of stability of AS to PW, whereas the onset of PW lies within the AS region leading to the bistable window, which is investigated in the following PW section.

\subsection*{(ii) Phase Wave state}
The PW state is characterized by a strong correlation between spatial and phase coordinates, typically takes the form 
\begin{align}
	x &\approx \pm \theta,
\end{align}

so that the collective distribution forms diagonal bands in the $(x,\theta)$ plane (see Fig. \ref{fig:1} (b)). The sign of the correlation ($+$ or $-$) is selected by memory effects of the hysteresis or the initial condition of the system. In terms of the order parameters $W_\pm = S_\pm e^{i\psi_\pm}$, the PW state is defined by one channel dominating while the other vanishes, i.e.

\begin{align}
	S_+ &> 0, \quad S_- \approx 0 
	\qquad \text{or} \qquad 
	S_- > 0, \quad S_+ \approx 0.
\end{align}

Thus, the PW corresponds to macroscopic coherence in either the $\epsilon$ or $\eta$ coordinate, while the conjugate coordinate remains incoherent. Dynamically, the PW shares phase boundaries with both the AS and BM states as seen in Fig. \ref{fig:4}, giving rise to bistable windows near both transitions.

To study the emergence of the hysteretic PW branch, we restrict to the PW sector
\begin{align}
S_+ \ge 0, \qquad S_- \approx 0, \nonumber
\end{align}
thereby reducing the self-consistency condition to a one-dimensional fixed point problem. To further limit the integration domain to the locked population, we introduce a Heaviside step function $H(x)$.  
After applying the locking condition (\ref*{eq:lock_condition}) and the tangent substitution $v=\tan s$, $\omega=\tan t$ \cite{ishitani2013transformations}, the self-consistency integral (\ref{eq:self_consistency}) becomes,

\begin{align}
	S_+ = \frac{1}{\pi^2} 
	\int_{-\tfrac{\pi}{2}}^{\tfrac{\pi}{2}} 
	\int_{-\tfrac{\pi}{2}}^{\tfrac{\pi}{2}} 
	\sqrt{\,1 - \left(\frac{A}{|\omega|(L/2)S_+}\right)^{2}} \notag \\
	\; H\!\big(|\omega|(L/2)S_+ - |A|\big)\,
	\mathrm{d}s\,\mathrm{d}t = \Phi(S_+,L).
	\label{eq:Phi_step}
\end{align}
where $A = \tan s + \tan t$ and $\omega = \tan t$, $L= J+K$ and $H$ enforces the locking condition: $H=1$ when $|A|\leq|\omega|J_+S_+$  for locked oscillators and otherwise $H=0$ for  drifting oscillators.  
Thus only the locked population contributes to the integral, while the rest of the population does not affect the order parameter.

For small $S_+$, we expand $\Phi(S_+,L)$ as
\[
\Phi(S_+,L) \approx \left.\frac{\partial \Phi}{\partial S_+}\right|_{S_+=0,L}\, S_+.
\]
If ${\partial \Phi}/{\partial S_+}(0,L) < 1$, the trivial solution $S_+=0$ corresponds to the stable AS state, while ${\partial \Phi}/{\partial S_+}(0,L) > 1$ indicates the birth of the PW branch. 

The integral $\Phi$ is evaluated numerically using Gauss–Kronrod quadrature \cite{Gautschi1988}, while the fixed point $S_+$ is obtained by solving
\begin{align}
	\Phi(S_+,L) - S_+ = 0
\end{align}

\begin{figure}[h!]
	\hspace*{-0.5cm}
	\includegraphics[width=8.5cm]{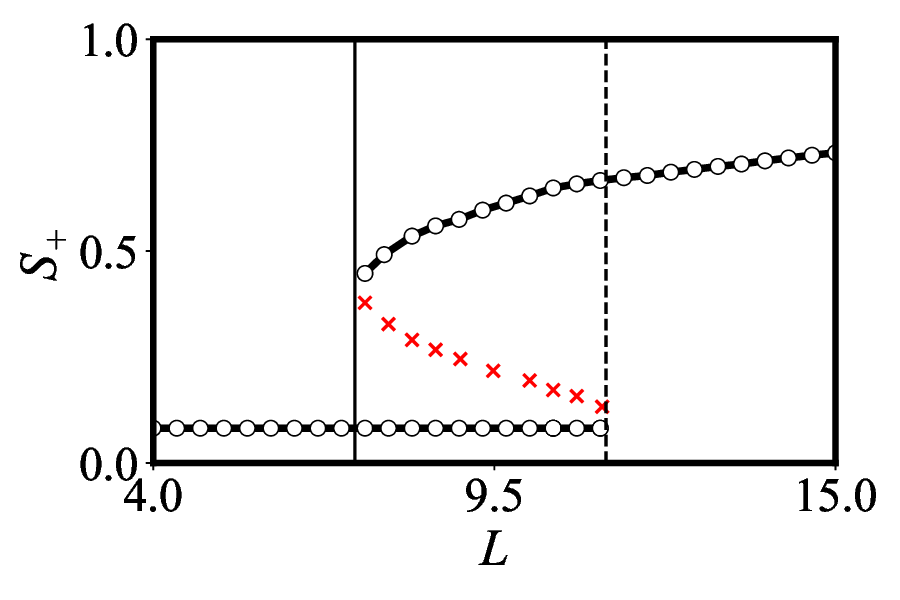}
	\caption{Numerical continuation of the PW branch in the $(S_+,L)$ plane. 
		The nonzero PW solution emerges through a saddle–node bifurcation at 
		$L_{c} \approx 7.2$(solid vertical line), corresponding to the backward PW$\to$AS boundary. Here, the circular markers represent stable solutions, whereas the red x markers are unstable solutions.
	}
	\label{fig:5}
\end{figure}

through Newton’s method. The corresponding continuation results are shown in Fig. \ref{fig:5} which traces both the zero and non-zero stable branches of $S_+$ along with an unstable branch corresponding to the bistable window. The solution terminates at a saddle–node bifurcation when the slope condition holds:
\begin{align}
	\left.\frac{\partial \Phi}{\partial S_+}\right|_{S_+,L} &= 1.
	\label{eq:PW_fold}
\end{align}

Thus, the forward (AS$\to$PW) and the backward (PW$\to$AS) boundaries were determined, by the fold condition (\ref{eq:PW_fold}) as depicted in Fig. \ref{fig:5}.

\medskip

We now proceed to study the loss of stability of the PW branch. The PW branch, defined by $S_+>0$ and $S_-\approx 0$, eventually loses stability within the BM region where both order parameters are nonzero. Hence, tracking the dynamics of $\chi$ reveals the onset of BM coherence and the corresponding loss of PW stability.

When $S_- > 0$ if we expand $\phi$ about PW locked solution, the $\chi$ dynamics (Eq.~(\ref{eq:chi_dyn})) reduce to
\begin{align}
	\dot{\chi} &= C - |\omega| D S_- \sin \chi,
\end{align}

with parameters
\begin{align}
	C &= B - \frac{J_-}{J_+} A,  \\
	D &= J_+ - \frac{J_-^2}{J_+} = \frac{JK}{J_+}. 
\end{align}

The locked solution then satisfies
\begin{align}
	\sin \chi^* &= \frac{C}{|\omega|DS_-}, \\[6pt]
	\cos \chi^* &= \sqrt{1 - \left(\frac{C}{|\omega|DS_-}\right)^2}.
\end{align}

Substituting into the definition of $S_-$ from integral (\ref{eq:self_consistency_general}) and averaging over the distributions gives
\begin{align}
	S_- &= \iint_{\text{locked}} g_\omega(\omega)\, g_v(A-\omega)\, 
	\cos\chi^* \, \mathrm{d}A\,\mathrm{d}\omega.
\end{align}
For a swarmalator to contribute to $S_-$ coherence, it must remain locked in both channels simultaneously:
\begin{align}
	|A| &\leq |\omega| J_+ S_+, \label{joint_ineq_1} 
\end{align}
\begin{align}
	|C| &\leq |\omega| |D| S_- . \label{joint_ineq_2}
\end{align}
The first inequality enforces PW coherence, while the second secures additional locking in the BM channel.  

For each $\omega$, the $\chi$ locking band in $A$ is defined by
\begin{align}
	C &= (1-\alpha)A - 2\omega, \\
	\alpha &= \tfrac{J_-}{J_+}. \nonumber
\end{align}

which centers the locked region at
\begin{align}
	A^* &= \frac{2\omega}{1-\alpha}.
\end{align}

with half-width
\begin{align}
	\Delta A &= \frac{|\omega|DS_-}{|1-\alpha|}.
\end{align}

Within this band, let's take $\sin\chi^* = X$ with $|X|\leq 1$, and the corresponding $\cos\chi^*$ contribution integrates to
\begin{align}
	\int_{-\Delta A}^{\Delta A} \cos \chi^* \,\mathrm{d}A 
	&= \frac{|\omega||D|S_-}{|1-\alpha|}
	\int_{-1}^{1} \sqrt{1-X^2}\,\mathrm{d}X \notag \\
	&= \frac{\pi}{2}\,\frac{|\omega||D|S_-}{|1-\alpha|}.
	\label{PW:lock_contribution}
\end{align}

By substituting the inner integral's contribution from (\ref{PW:lock_contribution}) and assuming the locking band in $A$ to be narrow, we approximate $g_v(A-\omega)$ by its value at the band center $A^*$, leading to
\begin{align}
	S_- \approx \int_{-\infty}^{\infty} 
	g_\omega(\omega)\, g_v(A^*-\omega)\,
	\left[ \frac{\pi}{2}\,\frac{|\omega||D|S_-}{|1-\alpha|} \right] \mathrm{d}\omega.
\end{align}

For unit-width Cauchy case, the self-consistency reduces to
\begin{align}
	\frac{\pi}{2}\,\frac{|D|}{|1-\alpha|}\,
	\frac{1}{\pi^2}\int_{-\infty}^\infty 
	\frac{|\omega|}{(1+\omega^2)\bigl(1+(J\omega/K)^2\bigr)}\,
	\mathrm{d}\omega &= 1.
\end{align}

Evaluating the integral yields the stability criterion
\begin{align}
	PW_c &= \frac{JK|K|}{\pi(J^2-K^2)} 
	\ln\!\left(\frac{|J|}{|K|}\right) = 1.
	\label{PW:loss}
\end{align}

which gives the condition for the loss of stability of the PW state.

\subsection*{(iii) Bi-strip mixed state}
\label{subsec:bistrip}

The BM state is a partially locked configuration of the swarmalators                                                                                                                                                                                                                                                                                                                                                                                                                                                                                                                                                                                                                                                                                                                                                                                                                                                                                                                                                                                                                                                                                                                                                                                                                                                                                                                                                                                                                                                                                                                                                                                                                                                                                                                                                                                                                                                                                                                                                                                                                                                                                                                                                                                                                                                                                                                                                                                                                                                                                                                                                                                                                                                                                                                                                                                  in which:
(i) two antipodal clusters appear in $(x,\theta)$ due to the $\pi$-shift symmetry
$(x,\theta)\mapsto(x+\pi,\theta\pm\pi)$, and (ii) each cluster splits into two thin, nearly parallel sub-strips (Fig. \ref{fig:1} (c)).
These sub-strips are populated by oscillators with $\omega>0$ and $\omega<0$, respectively; 

\begin{figure}[h!]
	\hspace*{-0.5cm}
	\includegraphics[width=7.0cm]{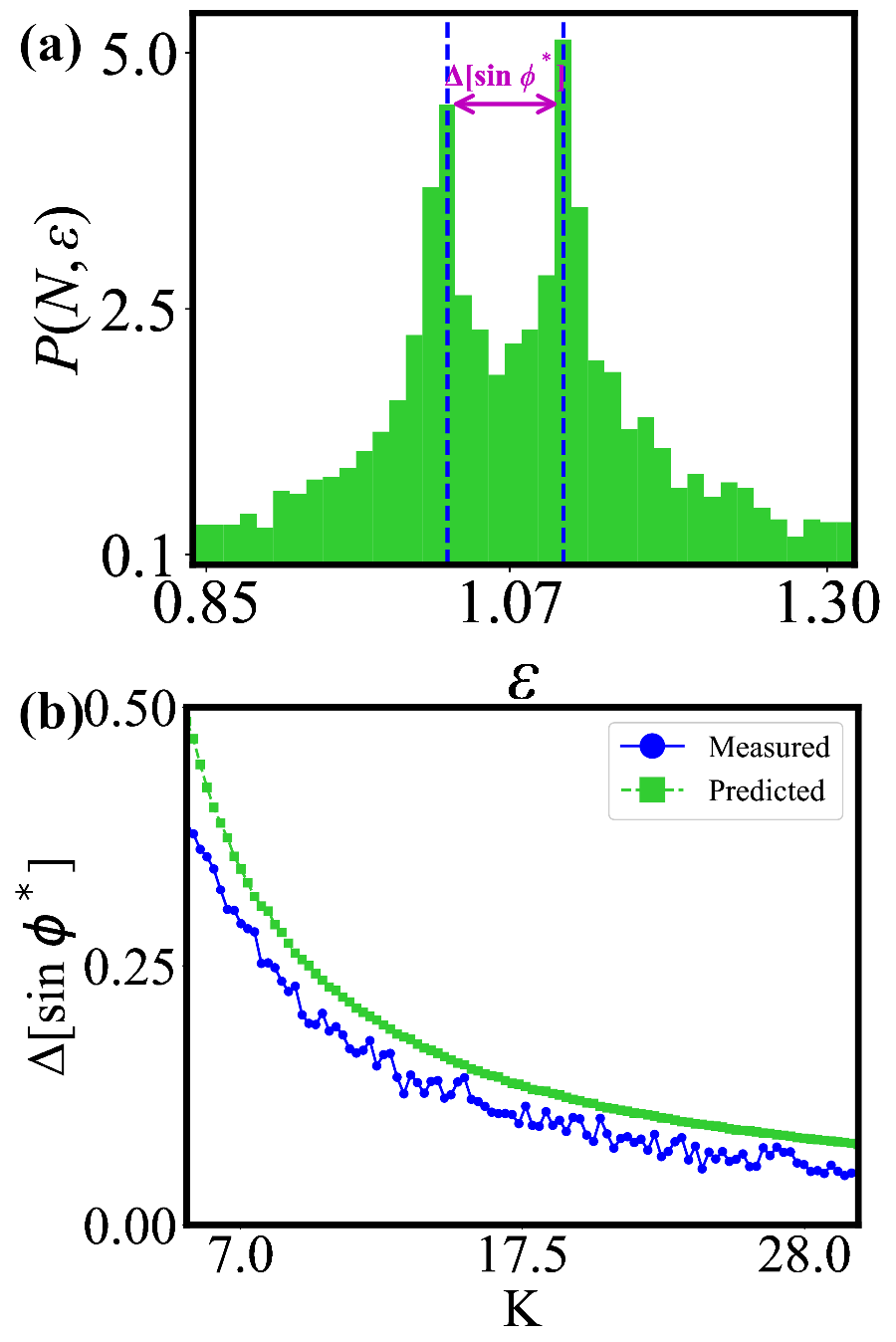}
	\caption{(a) Distribution of swarmalators in $\epsilon$ channel for parameters $J=12$ and $K=26$, (b) Separation $\Delta[\sin\phi^*]$ between substrips as a function of $K$, comparing theoretical prediction $\Delta = 2/(K S_+)$ with numerical measurements. }
	\label{fig:6}
\end{figure}

In BM state, the swarmalators are locked in both channels $\dot{\varepsilon}=\dot{\eta}=0$.
Then the Eqs. \eqref{eq:epsilon_dot}–\eqref{eq:eta_dot} solves exactly to

\begin{align}
	\sin\phi^* = \frac{1}{S_+}\!\left(\frac{v}{J|\omega|} + \frac{\operatorname{sgn}(\omega)}{K}\right),
	\\
	\sin\chi^* = \frac{1}{S_-}\!\left(\frac{v}{J|\omega|} - \frac{\operatorname{sgn}(\omega)}{K}\right).
\end{align}
The splitting of parallel strips is a direct consequence of the $\operatorname{sgn}(\omega)$ term in the locking
conditions. The discrete offset $\pm \frac{1}{K}\operatorname{sgn}(\omega_i)$ causes the locked equilibria to shift in opposite directions for $\omega_i>0$ and $\omega_i<0$, This shift splits each cluster into two parallel sub-strips, as seen in Fig. \ref{fig:2}. To find the separation between these sub-strips, we compare the fixed points for positive and negative natural frequencies:
This separation can be quantified as follows: 

\begin{align}
 \Delta [\sin \phi^*] 
 = \sin \phi^*_+ - \sin \phi^*_- 
 = \frac{2}{K S_+}.
 \label{strip:sep}
\end{align}
Here, $\sin\phi^*_{\pm}$ denote the locked solutions for $\pm \omega$. Building on this, we observed that the swarmalators exhibit a bimodal probability distribution in both the $\epsilon$ and $\eta$ channels. As illustrated in Fig.~\ref{fig:6}(a), we plotted the probability density $P(N,\epsilon)$ vs $\epsilon$ and it displays two peaks corresponding to $\omega > 0$ and $\omega < 0$, respectively. In this context, the sub-strip separation is calculated as the difference between the means of the two peaks in the bimodal distribution. A comparison is shown in Fig. \ref{fig:6} (b), where the theoretical prediction from (\ref{strip:sep}) is plotted alongside the measured separations from simulations for varying $K$, revealing that the inter-strip separation $\Delta [\sin \phi^*]$ decays with increased coupling strength $K$.

Next, we describe the transition to the BM branch. The BM branch begins with both $S_+>0$ and $S_->0$. This onset occurs through a saddle–node bifurcation, similar to the PW onset.  
To define this transition precisely, its onset can be determined from the coupled self-consistency relations
\begin{align}
 S_+ &= \iint_{\text{locked}} g_v(v)\,g_\omega(\omega)\,\cos\phi^*\,\mathrm{d}v\,\mathrm{d}\omega, \label{eq:BM_self_consistency_plus} \\
 S_- &= \iint_{\text{locked}} g_v(v)\,g_\omega(\omega)\,\cos\chi^*\,\mathrm{d}v\,\mathrm{d}\omega, \label{eq:BM_self_consistency_minus}
\end{align}
where the locked domains are defined by the joint inequalities defined in conditions (\ref{joint_ineq_1}), (\ref{joint_ineq_2}).

To evaluate these integrals and explore the onset conditions, we solved the equation $S=F(J,K;S)$ with $S=(S_+,S_-)$ using the same solver as mentioned in section III(B)(ii).  
In the backward continuation, the lower fold point indicates the onset of bistability between PW and BM. The numerically continued branch for $J=12$ is displayed in Fig.~\ref{fig:3}, where stable BM solutions are denoted by \textcolor{blue}{$\blacktriangle$} markers.

\begin{figure}[h!]
	\hspace*{-0.5cm}
	\includegraphics[width=8.5cm]{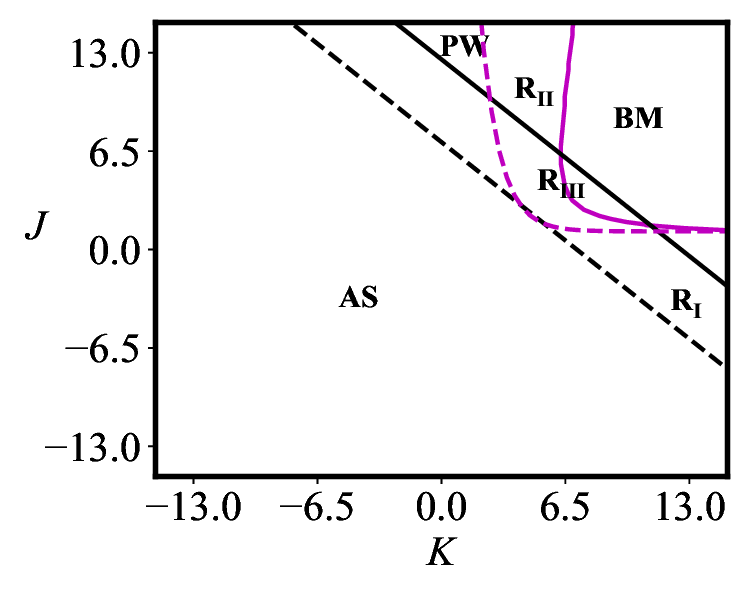}
	\caption{Analytic and Semi-analytic curves represented in $JK$ parameter space. Here, the solid magenta curve represents $PW_c$, solid black line represents $J_{+,c}$, dashed black line and dashed magenta curve represents the semi-analytical boundaries obtained for PW and BM onset respectively. The regions $R_{I},R_{II},R_{III}$ are depicted same as in the Fig. \ref{fig:4}}
	\label{fig:7}
\end{figure}

\section{Discussion}
\label{sec:discussion} 
Through extensive numerical simulations we have identified three distinct dynamical states in the frequency-weighted 1D swarmalator model. These include: (i) the AS state, where the swarmalators show no correlation in either $x$ or $\theta$; (ii) the PW state, where the population aligns along a diagonal correlation between $x$ and $\theta$; and (iii) the BM state, where two parallel sub-clusters appear within each antipodal cluster, separated by the sign of the natural frequency $\operatorname{sgn}(\omega)$. 

The BM state is reminiscent of the standing wave (SW) state in the Kuramoto model, where oscillators split into counter-rotating groups. However, a key distinction is that in the classical model SW states require a symmetric bimodal frequency distribution \cite{martens2009exact,pazo2009existence,crawford1994amplitude}. By contrast, in our case the BM state emerges even with a unimodal distribution, solely due to the frequency-weighted coupling. Moreover, in SW the order parameter traces a limit cycle \cite{xu2016synchronization}, leading to oscillatory nature while in our case the order parameters are stationary at around $S_{\pm} \approx 0.8$.

This type of frequency-based clustering has also been observed in the two-dimensional swarmalator model as ``bouncing clusters'', where spatial clusters separate according to the opposite signs of natural frequencies \cite{ceron2023diverse}. These observations demonstrate the robustness of the 1D toy model that, even under minimal modifications such as frequency-weighted coupling, reproduces the clustering seen in higher-dimensional systems. In the absence of frequency-weighted coupling, the counter-rotating clusters in BM state do not emerge, instead, only the mixed state with two anti-podal clusters reported in Ref. \cite{yoon2022sync} is observed.

The system also exhibits hysteretic transitions between states, caused by the saddle–node bifurcations associated with each boundary. From the self-consistency analysis, we find that the AS state loses stability at the forward threshold $J_{+,c}$, while the PW state loses stability at its critical boundary $PW_c$. The backward stability limits of the PW and BM states, corresponding to the lower fold points in the hysteretic window, were obtained by numerical continuation. The analytical thresholds and the numerically continued critical curves are summarized in Fig.~\ref{fig:7}.

Beyond the results reported here, several promising directions remain open. A natural extension is to study frequency-weighted swarmalators in two and three spatial dimensions. Another direction is to explore alternative heterogeneities, for example by combining frequency weighting with additional effects such as noise, delay, or pinning. The choice of frequency distribution also plays a key role, while Cauchy disorder is analytically convenient, it will be important to test how robust the observed hysteresis and clustered states are for Gaussian or heavy-tailed distributions. 

\begin{acknowledgments}
The work of R.G. and V.K.C. forms part of a research project sponsored by ANRF-DST-CRG Project Grant No. C.R.G./2023/003505. R.G. and V.K.C. thanks DST, New Delhi, for computational facilities under the DST-FIST programme (Grant No. SR/FST/PS-1/2020/135) to the Department of Physics.
\end{acknowledgments}

\section*{Data Availability}
The code that support the findings in this article are provided by the corresponding authors upon reasonable request.

\nocite{*}

\bibliographystyle{apsrev4-2} % Use a suitable style; this is good for revtex
\bibliography{references}     % Name of your .bib file (without .bib extension)

\end{document}